\documentclass[sigconf]{acmart}

\AtBeginDocument{%
  }

\copyrightyear{2026}
\acmYear{2026}
\setcopyright{cc}
\setcctype{by}
\acmConference[ACM Sustainability Week Companion '26]{ACM Sustainability Week 2026}{June 22--25, 2026}{Banff, AB, Canada}
\acmBooktitle{ACM Sustainability Week 2026 (ACM Sustainability Week Companion '26), June 22--25, 2026, Banff, AB, Canada}
\acmDOI{10.1145/3765611.3815137}
\acmISBN{979-8-4007-2199-1/2026/06}

\usepackage[acronym,toc]{glossaries}
\usepackage{amsmath}
\usepackage{booktabs}
\usepackage{graphicx}
\usepackage{subcaption}
\usepackage{algorithm}
\usepackage{algorithmic}
\usepackage{xcolor}
\usepackage{multirow}
\usepackage{balance}

\begin{document}

\title[FlowGuard: Identity-Independent Detection of Model Stealing]{FlowGuard: Flow Matching for Identity-Independent Detection of Data-Free Model Stealing Attacks on Energy System Intrusion Detection Systems}

\author{Maxime Schwarzer}
\affiliation{%
  \institution{CortAIx Labs, Thales Deutschland / Karlsruhe Institute of Technology}
  \country{Germany}
}
\email{maxime.schwarzer@thalesgroup.com}

\author{Laurin Holz}
\affiliation{%
  \institution{CortAIx Labs, Thales Deutschland}
  \country{Germany}}
\email{laurin.holz@thalesgroup.com}

\author{Tobias Huerten}
\affiliation{%
  \institution{CortAIx Labs, Thales Deutschland}
  \country{Germany}
}
\email{tobias.huerten@thalesgroup.com}

\author{Johannes Loevenich}
\affiliation{%
  \institution{CortAIx Labs, Thales Deutschland}
  \country{Germany}}
\email{johannes.loevenich@thalesgroup.com}

\author{Thies Moehlenhof}
\affiliation{%
 \institution{CortAIx Labs, Thales Deutschland}
 \country{Germany}}
\email{thies.moehlenhof@thalesgroup.com}


\author{Roberto Rigolin F. Lopes}
\affiliation{%
  \institution{CortAIx Labs, Thales Deutschland}
  \country{Germany}}
\email{roberto.rigolin@thalesgroup.com}

\author{Veit Hagenmeyer}
\affiliation{%
  \institution{Karlsruhe Institute of Technology}
  \country{Germany}}
\email{veit.hagenmeyer@kit.edu}

\renewcommand{\shortauthors}{Schwarzer et al.}
        \newacronym{ACD}{ACD}{Autonomous Cyber Defence}
\newacronym{IAI}{IAI}{Institute for Automation and Applied Informatics} 
\newacronym{BES}{BES}{battery energy storage} 
\newacronym{DER}{DER}{distributed energy resources} 
\newacronym{p.u}{$p.u$}{per unit} 
\newacronym{VSCs}{VSCs}{voltage source converters} 
\newacronym{DSO}{DSO}{Distribution System Operator} 
\newacronym{RMS}{RMS}{root mean square} 
\newacronym{AV}{AV}{Actual Value} 
\newacronym{TV}{TV}{Target Value} 
\newacronym{C4ISR}{C4ISR}{Command, Control, Communications, Computers, Intelligence, Surveillance, and Reconnaissance}
\newacronym{SGs}{SGs}{Smart Grids}
\newacronym{ML-based IDSs}{ML-based IDSs}{ Machine Learning-based Intrusion Detection Systems}
\newacronym{DNN}{DNN}{Deep Neural Network}
\newacronym{FDI}{FDI}{Feature Distortion Index}
\newacronym{ML}{ML}{Machine Learning}
\newacronym[plural=MEAs]{MEA}{MEA}{Model Extraction Attack}
\newacronym{MLSys}{ML-based systems}{machine learning-based systems}
\newacronym{MLaaS}{MLaaS}{ML-as-a-Service}
\newacronym{FGSM}{FGSM}{Fast Gradient Sign Method}
\newacronym{CWA}{C\&W attack}{Carlini \& Wagner attack}
\newacronym{PGD}{PGD}{Projected Gradient Descent}
\newacronym{ZOO}{ZOO}{Zeroth-Order Optimization}
\newacronym{SOTA}{SOTA}{state-of-the-art}
\newacronym{SDN}{SDN}{Software-Defined Networking}
\newacronym{DRL}{Deep RL}{Deep Reinforcement Learning}
\newacronym{GQM}{GQM}{Goal-Question-Metric}
\newacronym{AI}{AI}{Artificial intelligence}
\newacronym{MTS}{MTS}{Multivariate time series}
\newacronym{NLP}{NPL}{Natural Language Processing}
\newacronym{AI}{AI}{Artificial Intelligence}
\newacronym[plural=AICAs]{AICA}{AICA}{Autonomous Intelligent Cyber-defence Agent}
\newacronym{AGI}{AGI}{Artificial General Intelligence}
\newacronym{API}{API}{Application Programming Interface}
\newacronym{APIs}{APIs}{Application Programming Interfaces}
\newacronym[plural=APTs]{APT}{APT}{Advanced Persistent Threat}
\newacronym{AWS}{AWS}{Amazon Web Services}
\newacronym{AUC}{AUC}{Area Under the Curve}
\newacronym{DL}{DL}{Deep Learning}
\newacronym{YAML}{YAML}{YAML Ain’t Markup Language}
\newacronym[plural=EDFs]{EDF}{EDF}{Empirical Distribution Function}

\newacronym{SGD}{SGD}{Stochastic Gradient Descent}
\newacronym{GAN}{GAN}{Generative Adversarial Network}

\newacronym{IaC}{IaC}{Infrastructure as Code}

\newacronym[plural=IDSs]{IDS}{IDS}{Intrusion Detection Systems}
\newacronym{ID}{ID}{Intrusion Detection}
\newacronym{IPS}{IPS}{Intrusion Prevention System}

\newacronym{IP}{IP}{Intellectual Property}

\newacronym[plural=LLMs]{LLM}{LLM}{Large Language Model}
\newacronym{LSTM}{LSTM}{Long Short-Term Memory}
\newacronym[plural=SVMs]{SVM}{SVM}{Support Vector Machine}
\newacronym{MLP}{MLP}{Multilayer Perceptron}
\newacronym{ML}{ML}{Machine Learning}
\newacronym{SML}{SML}{Shallow Machine Learning}
\newacronym{NN}{NN}{Neural Network}

\newacronym{kNN}{kNN}{k-Nearest Neighbors}
\newacronym{RL}{RL}{Reinforcement Learning}
\newacronym{APT}{APT}{Advanced Persistent Threat}
\newacronym{APTs}{APTs}{advanced persistent threats}
\newacronym{SDD}{SDD}{Software Defined Defence}
\newacronym{SDN}{SDN}{Software Defined Networking}
\newacronym{APMSA}{APMSA}{Adversarial Perturbation Against Model Stealing Attacks}
\newacronym{MLIDS}{MLIDS}{Machine learning-based intrusion detection systems}
\newacronym{I-FGSM}{I-FGSM}{Iterative Fast Gradient Sign Method}
\newacronym{MI-FGSM}{MI-FGSM}{Momentum Iterative Fast Gradient Sign Method}
\newacronym{PRADA}{PRADA}{Protecting Against DNN Model Stealing Attacks}
\newacronym{ADAM}{ADAM}{Adaptive Moment Estimation}
\newacronym{RQs}{RQs}{Research Questions}
\newacronym{NATO}{NATO}{North Atlantic Treaty Organization}

\newacronym{CNN}{CNN}{Convolutional Neural Network}
\newacronym{FCN}{FCN}{Fully-Connected Feed Forward Neural Network}
\newacronym{LLMs}{LLMs}{large language models}
\newacronym{CV}{CV}{Computer Vision}
\newacronym{RAC}{RAC}{Relational Attention Corrector}
\newacronym{MMD}{MMD}{Maximum Mean Discrepancy}
\newacronym{KL divergence}{KL divergence}{Kullback-Leibler divergence}
\newacronym{PCA}{PCA}{Principal Component Analysis}

\newacronym{CTGAN}{CTGAN}{Conditional Generative Adversarial Network}
\newacronym{ERENO}{ERENO}{Efficacious Reproducer Engine for Network Operations}
\newacronym{RKHS}{RKHS}{Reproducing Kernel Hilbert Space}
\newacronym{SCA}{SCA}{Single Client Assumption}
\newacronym{TP}{TP}{True Positives}
\newacronym{FP}{FP}{False Positives}
\newacronym{TN}{TN}{True Negatives}
\newacronym{FN}{FN}{False Negatives}
\newacronym{ROC}{ROC}{Receiver Operating Characteristic}
\newacronym{TPR}{TPR}{True Positive Rate}
\newacronym{FPR}{FPR}{False Positive Rate}
\newacronym{AUC}{AUC}{Area Under the Curve}
\newglossaryentry{backpropagation}{name=Backpropagation, text=backpropagation, description={Propagating gradients obtained from an error function in backward direction through a network}}
\newglossaryentry{latex}{
    name=\LaTeX,
    description={A document preparation system}
}

\newglossaryentry{example}{
    name=Example,
    description={An example entry}
}

\newacronym{QUEEN}{QUEEN}{Query Unlearning}

\newacronym{SCADA}{SCADA}{Supervisory Control and Data Acquisition}
\newacronym[plural=CNFs]{CNF}{CNF}{Continuous Normalizing Flow}
\newacronym{OOD}{OOD}{out-of-distribution}

\begin{abstract}
Artificial Intelligence (AI)-based Intrusion Detection Systems (IDS) deployed in energy infrastructure are vulnerable to model theft attacks, which allow adversaries to create evasive traffic offline. Current defences against model extraction rely either on identity-bound query monitoring, which is ineffective against distributed attackers (Sybil), or on prediction poisoning through soft-label perturbation, which is inapplicable to hard-label IDS deployments. Therefore, we propose FlowGuard, an identity-independent defence based on flow matching that classifies incoming queries as out-of-distribution (OOD) prior to IDS processing. This approach exploits the fact that queries generated synthetically for data-free model stealing attacks occupy a lower-dimensional manifold than real network traffic. This results in measurably lower log-likelihoods when using a Continuous Normalizing Flow that has been trained on legitimate data. We evaluate our method against PRADA and FDINet using MAZE and DisGUIDE attacks in single-client and distributed (100-client Sybil) settings. While PRADA's detection rate dropped to 0\% when the distribution changed, our defence maintained a stable detection rate across both settings without relying on identity information. We discuss the scope and limitations of the approach, and outline potential applications to data-dependent attacks.
\end{abstract}

\ccsdesc[500]{Security and privacy~Software security engineering}
\ccsdesc[300]{Computing methodologies~Artificial intelligence}
\ccsdesc[300]{Computing methodologies~Neural networks}

\keywords{Model Extraction Attack, Intrusion Detection System, Flow Matching, Out-of-Distribution Detection, Sybil Attack, Critical Infrastructure Security}
\begin{teaserfigure}
  \includegraphics[width=\textwidth]{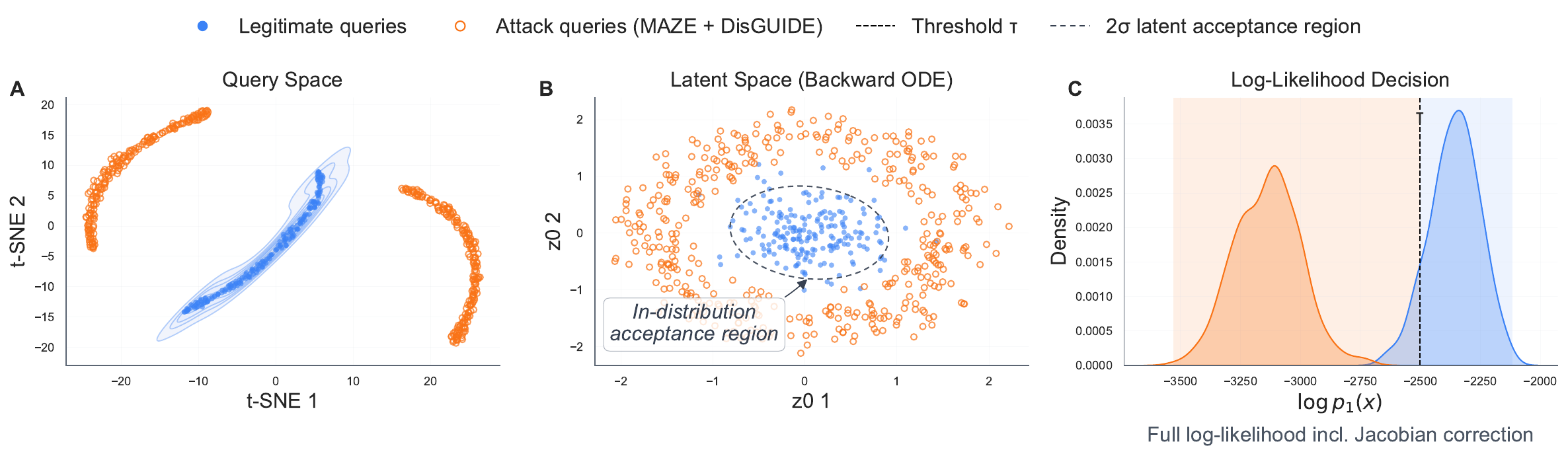}
    \caption{Flow Matching OOD detection. (A)~t-SNE of legitimate and synthetic attack queries in input space. (B)~Latent representations $z_0$ after backward ODE integration; attack queries fall outside the $2\sigma$ acceptance region. (C)~Log-likelihood distributions (Eq.~\ref{eq:density}) with decision threshold $\tau$. Each query is evaluated individually, independent of client identity.}
  \label{fig:teaser}
\end{teaserfigure}

        \maketitle

        \section{Introduction}
\label{sec:introduction}

\gls{AI}-based \gls{IDS} are increasingly deployed to protect critical energy infrastructure, including smart grids and \gls{SCADA} networks~\cite{khalaf2025development}. These systems use \glspl{DNN} to classify network traffic in real time. When such models are exposed through query interfaces, for instance, within a Security Operations Center or via internal \glspl{API} in a \gls{SDD} architecture, they become targets for \glspl{MEA}~\cite{tramer2016stealing, orekondy2018knockoff}.

In a \gls{MEA}, an adversary systematically queries the target model to train a functionally equivalent substitute. This substitute then serves as an offline testbed for crafting adversarial evasion traffic: network packets misclassified as benign by the original \gls{IDS}~\cite{papernot2017practical}. The compound threat of extraction followed by evasion is particularly severe in energy systems. Unlike traditional IT environments, where breaches primarily result in data loss, an undetected intrusion in a smart grid or \gls{SCADA} network can lead to direct physical consequences, including equipment damage, cascading failures, and widespread power outages.

Existing defenses against \glspl{MEA} fall into two categories, query detection and poisoning prediction, both with limitations in this setting. Query detection methods such as \gls{PRADA}~\cite{prada} and FDINet~\cite{fdinet} analyze incoming query patterns to identify anomalous behavior. However, these approaches operate on a per-identity basis, collecting statistics over queries from individual clients. Distributing queries across multiple identities via round-robin scheduling (a Sybil attack) reduces PRADA's detection to 0\%. Even global aggregation variants can be defeated through traffic mixing.

Prediction poisoning methods such as Adaptive Misinformation~\cite{adaptiveMisinformation} and MODELGUARD~\cite{tang2024modelguard} perturb the model's output probabilities to degrade the substitute model. These defenses require access to the full probability vector (soft labels). In practice, many \gls{IDS} deployments return only a binary hard label (attack/benign), rendering prediction poisoning inapplicable. Furthermore, even when applicable, model extraction remains feasible despite such perturbations~\cite{chandrasekaran2020exploring}.

While generative models and latent space representations have already been successfully applied to detect anomalies and model complex dynamics in energy systems~\cite{turowski2022enhancing, inn_paper1}, we now adapt this principle to secure these \gls{IDS} deployments against \glspl{MEA}. Specifically, we propose using a \gls{CNF} trained via Flow Matching~\cite{lipman2023flow, lipman2024flowmatchingguidecode} on the distribution of legitimate network traffic to classify incoming queries as in-distribution or \gls{OOD} before they reach the \gls{IDS}. FlowGuard operates on the query content rather than on query metadata or identity information, making it inherently resilient to Sybil attacks.

The key empirical observation is that data-free model stealing attacks (e.g., MAZE~\cite{maze}, DisGUIDE~\cite{rosenthal2023disguide}) generate synthetic queries that are useful for extracting decision-boundary information without necessarily matching the full training distribution. Under a density model trained on genuine traffic, these synthetic queries receive substantially lower log-likelihoods than legitimate queries. FlowGuard therefore treats low likelihood as the \gls{OOD} signal.

The motivation for applying Flow Matching comes from recent work on FlowPure~\cite{collaert2025flowpure}, which demonstrates that \glspl{CNF} trained with Conditional Flow Matching can effectively distinguish adversarial examples from clean inputs by measuring the magnitude of the learned velocity field. We adapt this principle to the distributed data-free model stealing detection problem.

The main contributions of this paper are as follows:
\begin{enumerate}
    \item We propose FlowGuard, an identity-independent defense against data-free model stealing attacks based on Flow Matching OOD detection.
    \item We evaluate the defense against MAZE and DisGUIDE attacks on CIFAR-10~\cite{cifar10}, a standard benchmark dataset for model stealing attacks, and compare against PRADA and FDINet.
    \item We discuss the scope of the approach, its current limitation to data-free attacks, and outline future extensions.
\end{enumerate}

This paper is organized as follows. Section~\ref{sec:related-work} provides the technical background on model stealing attacks, existing defenses, and the mechanics of normalizing flows. The proposed FlowGuard methodology is detailed in Section~\ref{sec:methodology}, explaining how Flow Matching is used to compute log-likelihoods for individual queries to detect synthetic attack data. Section~\ref{sec:threat-model} defines a threat model involving a black-box attacker utilizing data-free extraction methods. Section~\ref{sec:evaluation} discusses the experimental evaluation, benchmarking FlowGuard against PRADA and FDINet in both single-client and distributed Sybil attack settings. Finally, Section~\ref{sec:conclusion} summarizes the findings and suggests future work.

        \section{Background \& Related Work} \label{sec:related-work}

\subsection{Model Stealing Attacks (MEAs)}

A \gls{MEA} targets a victim model $f_V: \mathcal{X} \rightarrow \mathcal{Y}$ accessible through a query interface. The adversary trains a substitute model $f_S$ by querying $f_V$ with inputs $x \in \mathcal{X}$ and observing responses $\hat{y} \in \mathcal{Y}$. The goal is typically fidelity extraction, where $f_S$ replicates the decision boundary of $f_V$~\cite{tramer2016stealing}. Data-free attacks generate these queries synthetically: MAZE~\cite{maze} co-trains a generator with the substitute model, and DisGUIDE~\cite{rosenthal2023disguide} extends this with an ensemble-based disagreement and diversity loss. Both produce queries from noise without access to the original training data.


\subsection{Defenses Against Model Stealing}
\textbf{Query detection.} PRADA~\cite{prada} applies the Shapiro-Wilk test to the distribution of pairwise distances among queries from a single client, flagging deviations from normality. FDINet~\cite{fdinet}, the current state-of-the-art for detecting distributed attacks, computes a Feature Distortion Index from internal model activations to detect colluding clients. Both methods require accumulating a sufficient number of queries from individual identities before detection is possible.

\textbf{Prediction poisoning.} Methods such as Reverse Sigmoid~\cite{reverseSigmoid}, Prediction Poisoning/MAD~\cite{orekondy20prediction}, and MODELGUARD~\cite{tang2024modelguard} modify the returned probability vector to mislead the substitute model's training. These require soft-label access and do not prevent extraction under hard-label settings~\cite{chandrasekaran2020exploring}.


\subsection{Normalizing Flows and OOD Detection}
At their core, Normalizing Flows \cite{lipman2024flowmatchingguidecode} are bijective transformations that continuously map a complex, unknown data distribution (such as network traffic) into a simple, tractable base distribution (typically a standard Gaussian). This invertibility allows for the exact likelihood computation of new samples. Building on this, Flow Matching~\cite{lipman2023flow} trains \glspl{CNF} by regressing a neural network onto the conditional velocity field that transports samples between a source and target distribution. Unlike traditional normalizing flows, Flow Matching does not require expensive Jacobian computations during training, while still enabling exact density evaluation via the instantaneous change of variables formula at inference:
\begin{equation}
    \log p_1(x) = \log p_0(z_0) - \int_0^1 \text{tr}\left(\frac{\partial f_\theta}{\partial z}(z_t, t)\right) dt
    \label{eq:density}
\end{equation}
where $p_1$ denotes the data distribution, and a sample $x \sim p_1$ is interpreted as $z_1$ and mapped to a latent representation $z_0$ by integrating the learned ODE $\frac{dz}{dt} = f_\theta(z,t)$ backward from $t=1$ to $t=0$. The base distribution $p_0$ is typically chosen as a standard Gaussian.

The ability to compute exact log-likelihoods makes Flow Matching a candidate for \gls{OOD} detection, but likelihood direction must be calibrated for the concrete data and model. Deep generative models can assign higher likelihood to \gls{OOD} data than training data in certain cases (e.g., SVHN vs.\ CIFAR-10)~\cite{nalisnick2019deep}. In our setting, the calibrated signal is the lower tail of the likelihood distribution: legitimate validation queries define the accepted range, and synthetic extraction queries are flagged when their likelihood falls below that range.

        \section{FlowGuard: Flow Matching for OOD Detection}
\label{sec:methodology}

Our defense inserts a density-based filter between the query interface and the \gls{IDS}. The filter operates as follows:

\begin{enumerate}
    \item A Continuous Normalizing Flow $f_\theta$ is trained via Flow Matching on the same data distribution used to train the \gls{IDS}. The training follows the Conditional Flow Matching objective:
    \begin{equation}
        \mathcal{L}_{\text{CFM}}(\theta) = \mathbb{E}_{t, x_0, x_1}\left[\|f_\theta(x_t, t) - u_t(x_t \mid x_0, x_1)\|^2\right],
    \end{equation}
    where $x_0 \sim p_0$ (base Gaussian), $x_1 \sim p_{\text{data}}$, and $u_t$ is the target conditional vector field along straight paths from $x_0$ to $x_1$.

    \item When a query $x_q$ arrives at the \gls{IDS} interface, we treat it as a data-space sample at $t=1$ and integrate the learned ODE $\frac{dz}{dt}=f_\theta(z,t)$ backward to $t=0$ to obtain its latent representation $z_0$. The model then computes $\log p_1(x_q)$ via Equation~(\ref{eq:density}) as the base log-density $\log p_0(z_0)$ (with $p_0$ chosen as a standard Gaussian) corrected by the accumulated divergence term along the trajectory.

    \item If $\log p_1(x_q) < \tau$, the query is classified as \gls{OOD} and blocked. The threshold $\tau$ is a lower-tail threshold calibrated on a held-out validation set of legitimate queries.
\end{enumerate}

\subsection{Threat Model} \label{sec:threat-model}
We consider an adversary targeting an \gls{AI}-based \gls{IDS} deployed to protect energy infrastructure.

\textbf{Attacker capabilities.} The attacker has black-box query access to the \gls{IDS} and receives only hard labels (attack/benign). Given sufficient resources to coordinate queries across multiple identities (Sybil attack), the attacker uses data-free model stealing methods that generate queries from noise or co-trained generators without access to the original training data distribution.

\textbf{Attacker objective.} Extract a substitute model with high fidelity which enables crafting adversarial evasion traffic which bypasses the original \gls{IDS} decisions.

\textbf{Defender capabilities.} The defender has access to the legitimate training data distribution (or a representative sample) to train a density model. The defender can inspect each incoming query before it reaches the \gls{IDS}. The defender does not rely on any identity or session information.

\subsection{Why this works for data-free attacks.} 
Data-free model stealing methods generate queries through processes that do not have access to the full complexity of the real data distribution. MAZE \cite{maze} uses a generator co-trained with the substitute model, producing samples that approximate useful regions of the decision boundary rather than the full data manifold. DisGUIDE~\cite{rosenthal2023disguide} adds diversity through ensemble disagreement but still generates from noise. These objectives can produce queries that are informative to the victim classifier while remaining unlikely under a density model trained on legitimate data.

Unlike classifier confidence, the \gls{CNF} score measures compatibility with the legitimate data distribution. Data-free attacks optimize queries to extract target-model behavior, but they do not explicitly optimize likelihood under the defender's density model. As a result, their synthetic samples can be useful for extraction while still lying in low-density regions of the training distribution.

\subsection{Identity independence.} 
The defense evaluates each query individually based on its content. No identity information, session tracking, or query history is required. This makes the defense inherently resilient to Sybil attacks: even if an attacker distributes queries across thousands of identities, each individual query is still evaluated against the density model.

        \section{Evaluation} \label{sec:evaluation}
We evaluate whether content-based density filtering remains effective under distributed, identity-hiding model extraction, where identity-based detectors fail.

\subsection{Experimental Design}

Our evaluation follows the threat model from Section~\ref{sec:methodology}: a hard-label black-box adversary performing data-free extraction. We benchmark three query defenses under identical query budgets:
\begin{itemize}
    \item PRADA~\cite{prada}: Per-identity query distribution analysis (Shapiro-Wilk test on pairwise distances).
    \item FDINet~\cite{fdinet}: Feature Distortion Index computed from internal model activations.
    \item FlowGuard (ours): Log-likelihood threshold on the trained \gls{CNF}.
\end{itemize}

We test against two data-free attacks:
\begin{itemize}
    \item MAZE~\cite{maze}: Generator-based data-free attack with co-training.
    \item DisGUIDE~\cite{rosenthal2023disguide}: Ensemble-based generator with disagreement/diversity loss.
\end{itemize}

We evaluate two attacker regimes:
(1)~\textit{Single-client}, where all attack queries are issued under one identity, and
(2)~\textit{Distributed} (Sybil setting), where the same total budget is spread across 100 client identities.
This split tests whether a defense depends on identity-level temporal statistics (PRADA, FDINet) or on per-query content (FlowGuard).

\subsection{Protocol}

We use CIFAR-10 with a VGG16-BN victim model (85.03\% test accuracy) as the initial benchmark, following the standard evaluation protocol in model stealing research~\cite{prada, tang2024modelguard}. For each defense attack pair, we run one malicious and one benign episode:
(1)~200 attack queries generated by MAZE or DisGUIDE, and
(2)~200 clean reference queries sampled from legitimate data.
Metrics are computed at query level over the combined set.

\textbf{Flow model.} We train a \gls{CNF} using Meta's \texttt{flow\_matching} library for 1000 epochs on CIFAR-10. The velocity field is parameterized by a U-Net backbone and optimized with the Conditional Flow Matching objective (MSE between predicted and target velocity) using a Conditional Optimal Transport probability path from Gaussian noise to data. The threshold $\tau$ is selected empirically as a lower-tail cutoff from log-likelihood score distributions on validation data.

We report Detection Rate (equivalent to \gls{TPR}), \gls{FPR}, Precision, F1, Macro-F1, and \gls{ROC}--\gls{AUC}. This includes both threshold-dependent metrics (\gls{TPR}, \gls{FPR}, F1) and threshold-independent ranking quality (\gls{ROC}--\gls{AUC}).

\subsection{Results}

\begin{figure*}[t]
    \centering
    \includegraphics[width=\textwidth]{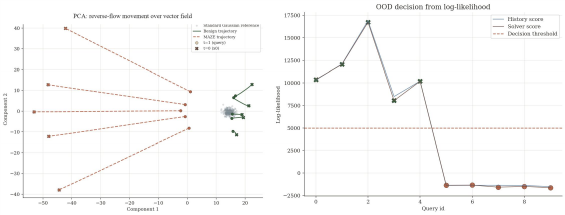}
    \caption{Representative MAZE deep dive for ten queries: five benign CIFAR-10 queries and five MAZE attack queries. The left panel shows the reverse-flow trajectory in a PCA projection of the learned flow dynamics, while the right panel shows the corresponding OOD decisions from log-likelihood. Benign queries remain above the calibrated lower-tail threshold, whereas MAZE queries fall below it and are blocked.}
    \label{fig:key-metrics}
\end{figure*}

Table~\ref{tab:detection-performance} summarizes the full aggregate detection experiment, while Fig.~\ref{fig:key-metrics} illustrates the corrected lower-tail score separation on a representative 10-query MAZE deep dive.

\begin{table*}[t]
    \centering
    \caption{Detection performance of query defenses against model-extraction attacks on CIFAR-10 (VGG16-BN target). Values are query-level metrics. Higher is better except false-positive rate (FPR), where lower is better.}
    \label{tab:detection-performance}
    \small
    \setlength{\tabcolsep}{4.0pt}
    \begin{tabular}{ll l c c c c c c c}
    \toprule
    Setting & Attack & Defense & Detection Rate & TPR & FPR$\downarrow$ & Precision & F1 & Macro-F1 & ROC-AUC \\
    \midrule
    \multirow{6}{*}{Single-client}
    & MAZE & FDINet & 0.545 & 0.545 & 0.530 & 0.507 & 0.525 & 0.507 & 0.488 \\
    & MAZE & PRADA & 0.840 & 0.840 & \textbf{0.000} & \textbf{1.000} & \textbf{0.913} & \textbf{0.919} & 0.920 \\
    & MAZE & FlowGuard & \textbf{0.965} & \textbf{0.965} & 0.170 & 0.850 & 0.904 & 0.897 & \textbf{0.921} \\
    & DisGUIDE & FDINet & \textbf{1.000} & \textbf{1.000} & 0.530 & 0.644 & 0.784 & 0.712 & 0.988 \\
    & DisGUIDE & PRADA & 0.833 & 0.833 & \textbf{0.000} & \textbf{1.000} & 0.909 & \textbf{0.918} & 0.917 \\
    & DisGUIDE & FlowGuard & \textbf{1.000} & \textbf{1.000} & 0.170 & 0.850 & \textbf{0.919} & 0.913 & \textbf{1.000} \\
    \midrule
    \multirow{6}{*}{Distributed (100 clients)}
    & MAZE & FDINet & 0.500 & 0.500 & 0.530 & 0.485 & 0.493 & 0.485 & 0.473 \\
    & MAZE & PRADA & 0.000 & 0.000 & \textbf{0.000} & 0.000 & 0.000 & 0.333 & 0.500 \\
    & MAZE & FlowGuard & \textbf{0.965} & \textbf{0.965} & 0.170 & \textbf{0.850} & \textbf{0.904} & \textbf{0.897} & \textbf{0.922} \\
    & DisGUIDE & FDINet & \textbf{1.000} & \textbf{1.000} & 0.530 & 0.644 & 0.784 & 0.712 & 0.989 \\
    & DisGUIDE & PRADA & 0.000 & 0.000 & \textbf{0.000} & 0.000 & 0.000 & 0.333 & 0.500 \\
    & DisGUIDE & FlowGuard & \textbf{1.000} & \textbf{1.000} & 0.170 & \textbf{0.850} & \textbf{0.919} & \textbf{0.913} & \textbf{1.000} \\
    \bottomrule
    \end{tabular}
    \vspace{2pt}
    \\
    {\footnotesize\emph{Note.} Bold values highlight best-performing entries for key decision metrics within each attack block.}
\end{table*}

\paragraph{Corrected likelihood direction.} The MAZE deep dive confirms that FlowGuard blocks low-likelihood queries. In the 10-query example shown in Fig.~\ref{fig:key-metrics}, benign CIFAR-10 queries receive scores between 8478.75 and 16885.64, while MAZE queries receive scores between -1504.09 and -1349.31. With $\tau=5000$, all benign queries remain above the threshold and all MAZE queries fall below it. This corrected direction is the basis for the aggregate metrics in Table~\ref{tab:detection-performance}.

\paragraph{Single-client setting.} Both PRADA and FlowGuard detect data-free attacks, but with different trade-offs. For MAZE, PRADA reaches \gls{TPR}~0.840, \gls{FPR}~0.000, and F1~0.913, while FlowGuard achieves higher recall (\gls{TPR}~0.965) at moderate \gls{FPR}~(0.170), yielding F1~0.904. For DisGUIDE, FlowGuard reaches \gls{TPR}~1.000 and ROC-AUC~1.000 with F1~0.919, whereas PRADA reaches TPR~0.833 and F1~0.909. FDINet shows comparable detection performance for DisGUIDE (F1~0.784) but fails to detect MAZE queries (\gls{ROC}--\gls{AUC}~0.488).

\paragraph{Distributed setting (100 clients).} PRADA collapses to no detection at all (\gls{TPR}~0.000, F1~0.000, \gls{ROC}--\gls{AUC}~0.500) for both attacks. This is expected: distributing queries across identities prevents sufficient per-client accumulation for the statistical test.

FlowGuard remains stable across both settings. For MAZE it keeps \gls{TPR}~0.965 and F1~0.904 (\gls{ROC}--\gls{AUC}~0.922); for DisGUIDE it keeps \gls{TPR}~1.000 and F1~0.919 (\gls{ROC}--\gls{AUC}~1.000). The per-query density evaluation is unaffected by how queries are distributed across identities.

FDINet shows mixed results. It detects DisGUIDE queries with high sensitivity (\gls{TPR}~1.000), but at high false-positive rate (\gls{FPR}~0.530), reducing precision. For MAZE under distributed querying, its \gls{TPR} drops to 0.500 with F1~0.493, indicating lower robustness than FlowGuard in this setup.

\subsection{Interpretation}

The results confirm that identity-independent, content-based detection is crucial for securing models against distributed extraction. PRADA, which relies on per-client statistical assumptions,  breaks down under Sybil coordination. FDINet, while designed to handle distributed attacks by analyzing internal feature distortions, faces a different limitation. Although its performance does not degrade under Sybil coordination, it struggles with an excessively high false-positive rate (\gls{FPR} 0.530) and fails to reliably detect queries generated by MAZE (\gls{ROC}--\gls{AUC} dropping to 0.473). This indicates a lack of robustness across different data-free generators. FlowGuard demonstrates that density-based filtering overcomes both challenges. Because it evaluates each query's log-likelihood independently and blocks lower-tail outliers, it is inherently immune to Sybil partitioning (unlike PRADA). Furthermore, by anchoring the decision boundary to the legitimate data distribution rather than relying on internal model activations, it maintains high detection rates across different attack generators.

The difference between \gls{ROC}--\gls{AUC} and fixed-threshold metrics is worth noting. High \gls{ROC}--\gls{AUC} indicates good ranking of benign vs.\ malicious queries across all thresholds, while \gls{FPR} and F1 depend on the chosen operating point $\tau$. A defense with near-perfect \gls{ROC}--\gls{AUC} can still exhibit non-zero false positives at a specific deployment threshold.

\subsection{Limitations}

The evaluation is limited to one dataset (CIFAR-10), one victim architecture, and one run per condition (no confidence intervals). The attack set covers data-free extraction only (MAZE, DisGUIDE). These results should be interpreted as evidence for the effectiveness against distributed data-free model stealing, not as a universal guarantee across all extraction families or data modalities. In particular, an adaptive attacker that explicitly optimizes against the defender's density model could reduce the observed likelihood gap. Extending to \gls{IDS}-specific datasets (e.g., ERENO IEC~61850), repeated seeds, and adaptive attackers is necessary future work.

        \section{Conclusion \& Future Work}
\label{sec:conclusion}

We present FlowGuard, a defense against data-free model stealing attacks based on Flow Matching \gls{OOD} detection. The approach classifies incoming queries by computing their log-likelihood under a \gls{CNF} trained on the legitimate data distribution. In our experiments, synthetic queries from data-free attacks receive consistently lower likelihoods than legitimate queries and can be reliably detected with a calibrated lower-tail threshold.

Our evaluation on CIFAR-10 shows that while PRADA's detection drops to 0\% under distributed querying (100 clients), FlowGuard maintains stable detection (\gls{TPR}~0.965--1.000) regardless of how queries are distributed. FDINet shows mixed behavior with high false-positive rates in some configurations. The per-query nature of the density filter makes it inherently resilient to Sybil attacks.

The current scope is limited to data-free attacks. For future work, we plan to:
\begin{itemize}
    \item Extend the evaluation to realistic \gls{IDS} datasets (ERENO IEC 61850, CIC-IDS) and additional defenses (D-ADD~\cite{dadd}, MeCo~\cite{wang2023defending}).
    \item Investigate the combination of FlowGuard with techniques used in adversarial purification. FlowPure~\cite{collaert2025flowpure} demonstrates that \glspl{CNF} can also detect adversarial examples through velocity field magnitude at $t=0$. Combining density-based detection of extraction queries with velocity-based detection of adversarial evasion traffic could provide a unified defense against the full model-stealing-then-evasion attack chain.
    \item Investigate whether more sophisticated generators (e.g., diffusion-based) can produce queries that overcome the density filter and develop corresponding countermeasures.
\end{itemize}

        \balance
        \bibliographystyle{ACM-Reference-Format}
        \bibliography{sources}
        \balance

\end{document}